# Frequency demodulation with magnetoelectric coreshells: A novel approach to enhanced bio-stimulation

Ram Prasadh Narayanan
ram.p.narayanan@ntnu.no
Norwegian University of Science and Technology
Trondheim, Norway

Ali Khaleghi
Norwegian University of Science and Technology
Intervention Center, Oslo University Hospital
Trondheim, Oslo, Norway

Ilangko Balasingham
Norwegian University of Science and Technology
Intervention Center, Oslo University Hospital
Trondheim, Oslo, Norway

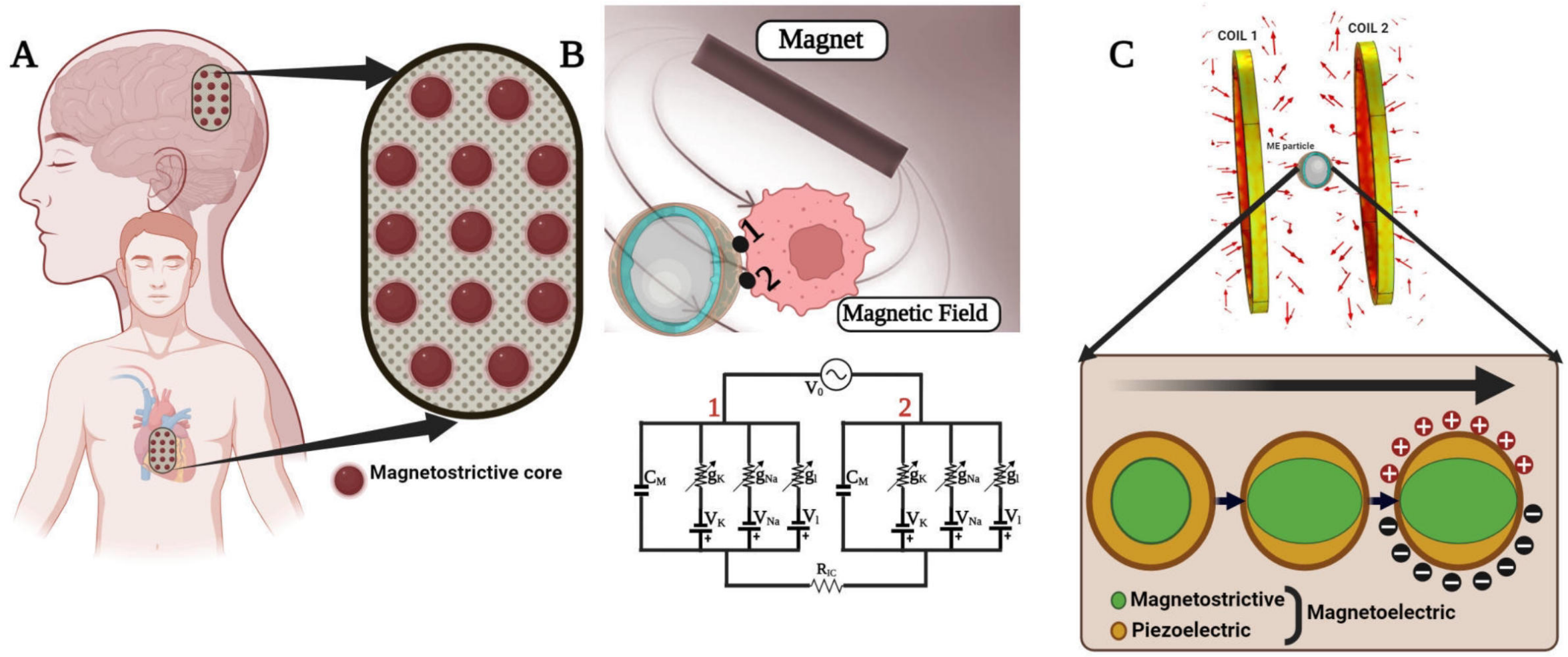

**Figure 1: (A) Array of Magnetostrictive cores fabricated as a single patch device could be used for non–invasive magnetic field based stimulation. (B) A minimally invasive ME coreshell and its interaction with a single cell is shown with the equivalent circuit model. (C) The action of magnetostriction and the generation of electrical potential through temporal interference frequency generated from two coils is also shown.**

## ABSTRACT

Magnetoelectric (ME) coreshell devices have interesting applications in biomedical technologies including biosensing, and communication, due to their strong inter–coupling between the magnetostrictive core and the piezoelectric shell. This property could be utilized for specific applications in localized bio–stimulation of cells. This paper provides a conceptual proof of using the non–linear property of ME coreshell devices for frequency demodulation and stimulation. We use the Multiphysics simulation approach, wherein the ME coreshell was biased with a DC magnetic field and perturbed through dual coil alternating magnetic fields. The combined effect of the ME non–linear magnetostriction and the dual coil perturbation resulted in the demodulation of the interference frequency component, that induced equivalent 'electrical hotspots' on the piezoelectric shell. We provide a cross model verification, where the generated electrical current density on the piezo shell was provided as an input to a Hodgkin–Huxley (HH) neural cell model to actively induce membrane potentials on the cell. Future applications as a standalone, battery- and electronics-free, controllable, multi–functional and localized coreshells for targeted drug delivery, backscatter communication and bio–stimulation, is envisioned.

# 1 INTRODUCTION

Electrophysiological stimulation is a method of using electrical currents to perform precise and localized stimulation at different levels, including organ, tissue, and cellular levels [16, 24, 25]. It has been extensively researched and used for theronostic and neuro–stimulation, with potential applications in the treatment of various conditions, such as depression, migraine, paralysis treatment and rehabilitation, Parkinson's, and chronic pain relief [12]. Non–contact and high resolution methods of stimulation include biological pacemakers based on gene modified cells [8, 22], optogenetic [19], chemogenetic [6], and magnetogenetic stimulation [5]. Most of these techniques require either the mediation of optical or magnetic nanoparticles that act as either down conversion agents for spectrum conversion and eventual stimulation using the localized source [4, 7] or the transduction of optical or RF absorbed power to thermal energy for subsequent stimulation [9]. For other non–contact stimulation techniques including Transcranial Magnetic (TMS), and Alternating Current Stimulation (TACS), the resolution of the techniques highly dependent on various factors including the coil design, directionality, electrical dissipation in the physiological medium, which also affects the penetration depth of such treatments [3, 15, 23]. Microelectrode arrays with electrical stimulation are used in brain-machine interfaces (BMIs) for medical purposes and brain rehabilitation [18]. However, miniaturization and power management of these devices are challenging, and wireless power transfer using inductive coupling and radio frequency energy transfer is used for such minimally invasive electrodes.

In this paper, we discuss the concept of using localized agents of battery- and electronics- free ME coreshell microdevices as frequency demodulators and their use case scenario in remote electrical bio–stimulaion. Magnetoelectric (ME) coreshell devices are hybrid composites with a magnetostrictive core and a piezoelectric shell. The strong inter–coupling between these two layers, defined by the ME coupling coefficient, $\alpha_{ME}$ could be exploited to induce magnetic fields through applied electrical currents (Villari effect) or vice–versa (Joule effect). This property opens a gateway to exciting applications with ME coreshells in medical technologies. The teaser in Figure 1 illustrates the use of ME coreshell devices as contact and injectable devices. ME coreshells effectively couple a magnetization induced deformation on the magnetostrictive core to an electric field on the piezoelectric shell. Therefore, we show that through precise coil placement and control of the external magnetic fields, the non–linear property of the ME coreshell could be exploited to demodulate the frequency components of the applied magnetic fields, $f_1$, $f_2$, and the difference frequency component, $f_1 - -f_2$. The non-linear magnetostriction property of ME coreshells can be used to demodulate the interference frequency component from two high-frequency signals, which can then be utilized as a localized source for low-frequency stimulation at the tissue level. Additionally, the highly non-linear and irregular deformation of the coreshells at their mechanical resonances induces irregular electrical hotspots, without the use of any implantable electronics. The proposed method has significant advantages over conventional bio–stimulation techniques, including superior spatial resolution and energy efficiency, resulting in reduced external unit size and complexity. We calculated the current density developed on the coreshell surface and provided it as an input to the Hodgkin–Huxley (HH) cell model. The equivalent cell model along with the ME device connected to its extracellular surface, is shown in Figure 1B.

# 2 SYSTEM MODEL

## 2.1 Simulation setup

A single ME coreshell geometry was simulated inside a physiological medium. Material properties of muscle, fat and skin layers were added from the references here [10, 17]. We used COMSOL Multiphysics, a finite element method (FEM)–based simulation software [2], and incorporated the Solid Mechanics *(sm)*, Electrostatics *(es)* and Magnetic Fields *(mf)* modules coupled to form Magnetostriction and the Piezoelectric Multiphysics. Figure 2 depicts a 2D schematic of the simulated COMSOL geometry, showing the two remote coils (numbered as '1' and '2'), the physiological medium and the ME coreshell device, as shown in the inset of Figure 2A. The application of the external DC magnetic bias and the alternating magnetic fields from the two symmetric coils, induces a magnetic field on the magnetostrictive core, subsequently inducing a piezoelectric coupling to produce an electric potential difference on the piezoshell, as shown in Figure 2B. We used the commercially available 2628 MB MetGlas as the magnetostrictive spherical core due to its high coupling coefficient [1, 14] and commercially available Aluminum Nitride (AlN) as the piezoelectric shell. We restricted the size of the simulated magnetostrictive core to 100 $\mu$m, so that the overall size of the ME coreshell device was in the sub–$\mu$m range, thus enabling it to be used as an injectable material. The external magnetic DC magnetic bias was set at 293 mT (it was estimated through supporting simulations, as the requirement to achieve maximum magnetostriction) and applied in the axial direction of the magnetostrictive core. The coreshell was placed between the two symmetrically positioned coils driven by a constant current source of 0.1A. The coils produced AC magnetic field perturbation at frequencies of 126 and 188 MHz. We show that the non–linear behavior of the coreshell generates the MTI at the difference frequency ($f_1 - -f_2$). Subsequently, the MTI resultant AC magnetic field at 62 MHz, which is the difference frequency of the two applied high frequency fields, was used to induce magnetostriction in the core.

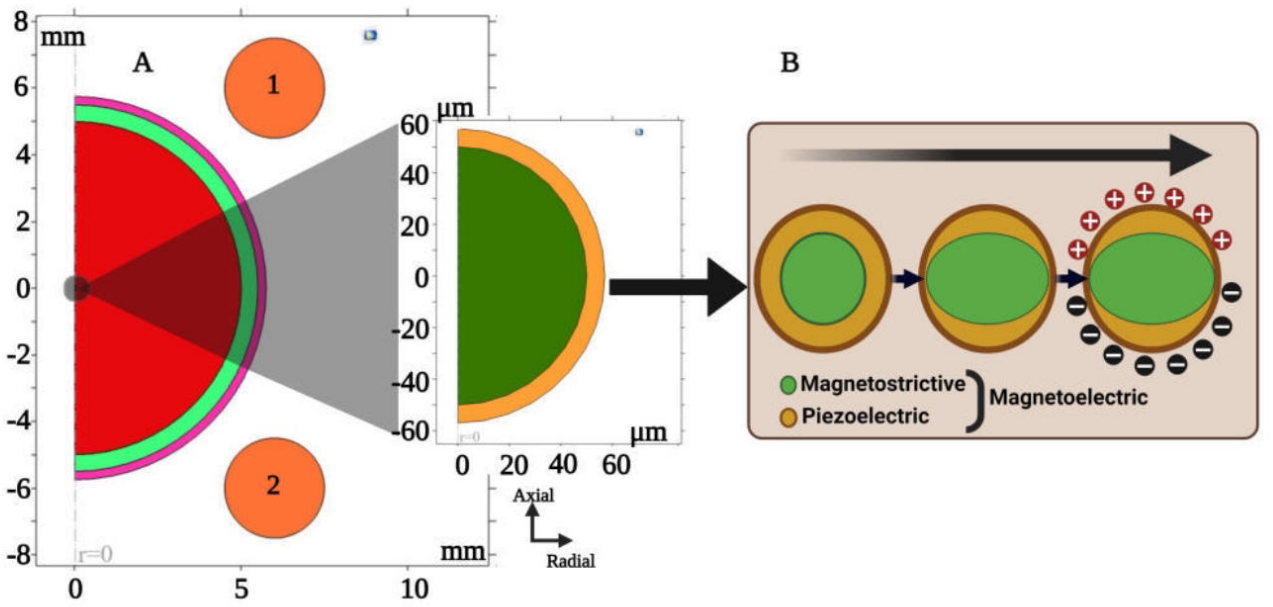

**Figure 2: (A) Simulation setup of the COMSOL geometry showing the coils, muscle, fat and skin layers (in bright orange, red, green and pink, respectively) and the ME device. (B) The MTI frequency, which is the difference frequencies of $f1$ and $f2$, from coils 1 and 2 respectively, is used to induce magnetostriction in the core, that results in the surface voltage. The geometry shown is axis symmetric. Hence only half of the model geometry is simulated.**

## 2.2 Theory of Magnetostriction, magnetoelectric behaviour and constitutive equations

Magnetostriction is the strain observed in ferromagnetic materials due to applied magnetic fields. The strain is a result of a strong orientation and weak perturbation (in that order) of the electron spin in the ferromagnetic material, resulting in the lattice deformation of the ferromagnetic structures [20, 27]. In other words, magnetostriction is caused due to the interaction of magnetic and elastic forces, thereby considered as a useful property for energy conversion between magnetic and piezoelectric domains, defined by the Joule and Villari effect. The governing equations for a ME device could be expresses as [26]

$$\sigma = \mathbf{C}\epsilon - \mathbf{e}^T E - \mathbf{q}^T H$$
$$D = \mathbf{e}\epsilon + \mathbf{k}E + \alpha H$$
$$B = \mathbf{q}\epsilon + \alpha E + \mu H$$

where, **C** is the elastic stiffness tensor, **e** and **q** are the piezoelectric and piezomagnetic constant tensors respectively, **k** and $\mu$ are the electric permittivity and the magnetic permeability tensors, $\alpha$ is the magnetoelectric coefficient tensor, $\epsilon$ and $\sigma$ are the mechanical strain and stress respectively, $E$ and $D$ are the electrical field and displacement respectively, and, $H$ and $B$ are the magnetic field and flux density respectively. COMSOL Multiphysics was used to simulate a strain–mediated ME effect on a coreshell, represented by two concentric spheres, where the inner and outer layer spheres were the Magnetostrictive and the Piezoelectric layers, respectively. Multiphysics entities, **Piezoelectric effect** and **Magnetostriction** were included with the coupling of **Solid Mechanics, Electrostatics and Magnetic Fields** modules. The magnetic field modeling was performed based on the constitutive B–H relation, set for all the domains (except the magnetostrictive device) (Boundary condition: Ampere's Law),

$$\vec{\mathbf{B}} = \mu_0 \mu_r \vec{\mathbf{H}}$$
$$\nabla \cdot \vec{\mathbf{H}} = \vec{\mathbf{J}}$$

and for the magnetostrictive core (*Boundary condition: Ampere's Law, Magnetostrictive*),

$$\vec{B} = \mu_0 [\vec{H} + \vec{M}(\vec{H}, S_{mech}) + \vec{M_r}]$$

where $\boldsymbol{B}$ is the magnetic flux density, $\boldsymbol{H}$ is the magnetic field intensity, $\boldsymbol{J}$ is the volumetric current density, $\mu_0$ is the permeability of free space, $\mu_r$ is the magnetic relative permeability, M is the magnetization, $S_{mech}$ is the stress tensor, and $M_r$ is the remanent magnetization, which is set as zero. The magnetostrictive stress is modeled as a non–linear isotropic entity given as

$$\epsilon_{me} = \frac{3}{2} \frac{\lambda_s}{M_s^2} dev(\vec{M} \otimes \vec{M})$$

where, $\lambda_s$ is the saturation magnetostriction, $M_s$ is the saturation magnetization of the magnetostrictive shell layer. The magnetization M, is given as a function of effective mahnetic field intensity $H_{eff}$ and $M_s$ as,

$$\vec{M} = M_s \quad L\left(|H_{eff}|\right) \frac{H_{eff}}{|H_{eff}|}$$

$$\vec{H_{eff}} = \vec{H} + \frac{3\lambda_s}{\lambda_0 M_s^2} S \ \vec{M}$$

where, L is the Langevin function and S is the stress tensor. For the Piezoelectric shell, The *Charge Conservation* boundary condition governs the operating Piezo shell given as,

$$\nabla \cdot \vec{\mathbf{F}} = \rho_v$$

where, $F$ is the electric flux density, and $\rho_v$ is the volume charge density.

## 2.3 Magnetic Temporal Interference (MTI)

In conventional MTI, the amplitude wave of the difference signal is generated by a linear correlation of two different applied frequencies, resulting in a beating envelope [15]. However, in ME mediated MTI based bio–stimulation technique, the magnetic field component's demodulation at the difference frequency is the output of the non–linear multiplication of two different applied frequencies. This is due to the elasto–magnetic effect on the magnetostrictive material, where the action of electromagnetic force on the magnetostrictive core is a function of the applied magnetic field, flux density, and generated stress tensor in the magnetostrictive domain. The force calculation on the magnetostrictive core is performed as an integral of the Maxwell stress tensor on the set domain surface, as follows:

$$n_1 T_1 = \frac{-1}{2} n_1 (H.B) + (n_1.H) B^T \tag{1}$$

where $n_1$ is the surface normal (1–by–3) vector passing outward to the surface, $T_1$ is the Maxwell stress tensor value in the magnetostrictive domain, and $H$ and $B$ are the magnetic field and flux density, respectively, acting on the domain surface. Figure 3 shows the magnetic field induced in the core at different time instances and the resultant deformation of the core due to the induced magnetic field. The non–linear hysteresis loop of the magnetostrictive core's B–H ratio and tensor results in non–linear "electrical hotspots" on the piezo shell. The contraction and expansion of the core structure could be interchanged by applying DC magnetic bias either in the axial or radial direction. In our simulation, the bias was applied in the axial direction, and for varying time instances, the contraction was found to be higher in the radial direction with subsequent expansion of the core in the axial direction.

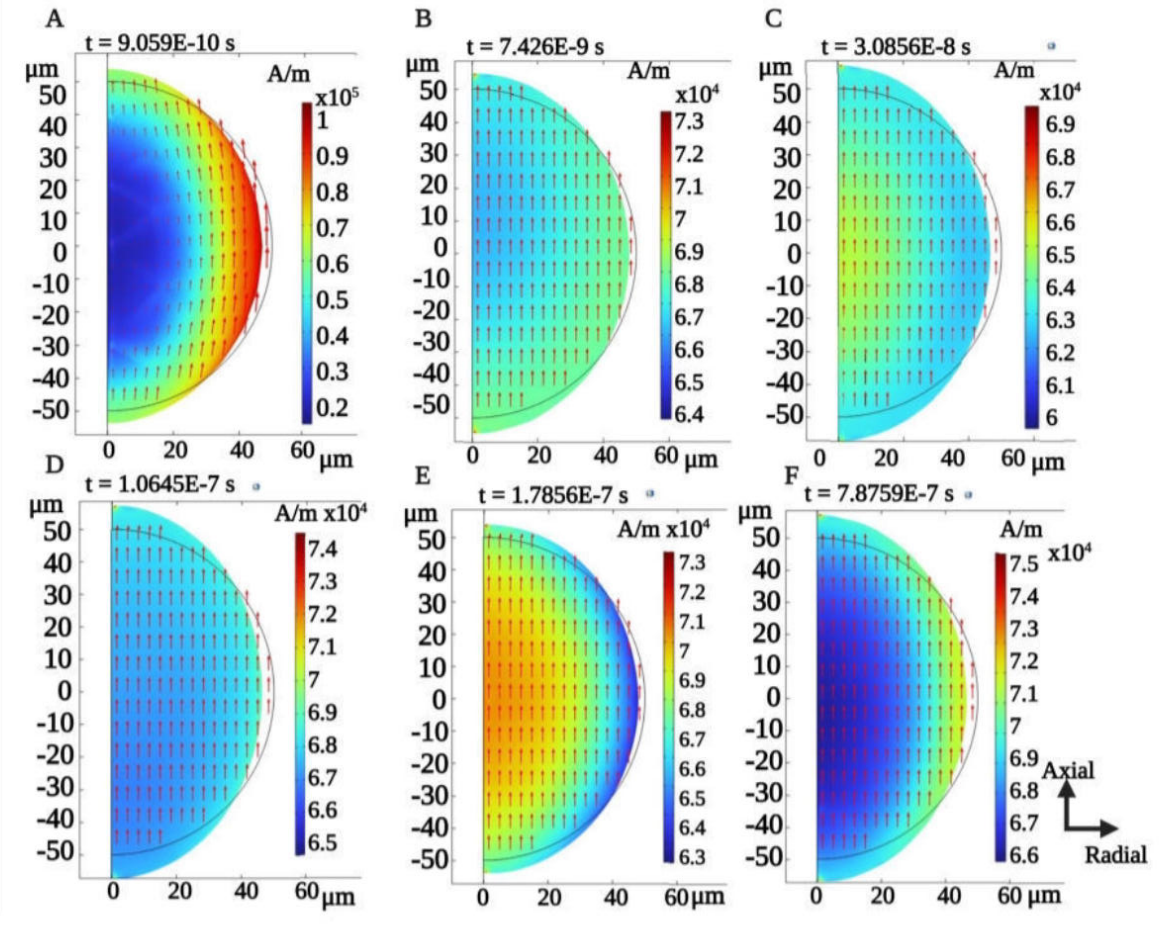

**Figure 3: Normalized magnetic field induced on the Magnetostrictive core for an applied DC magnetic bias of 293 mT (A–F). The deformation of the core due to the magnetic elastic energy transfer is also shown. At an applied magnetic DC magnetic bias of 293 mT, the surface average of the induced magnetic field was calculated to be** $\approx 66.7\ kA/m$

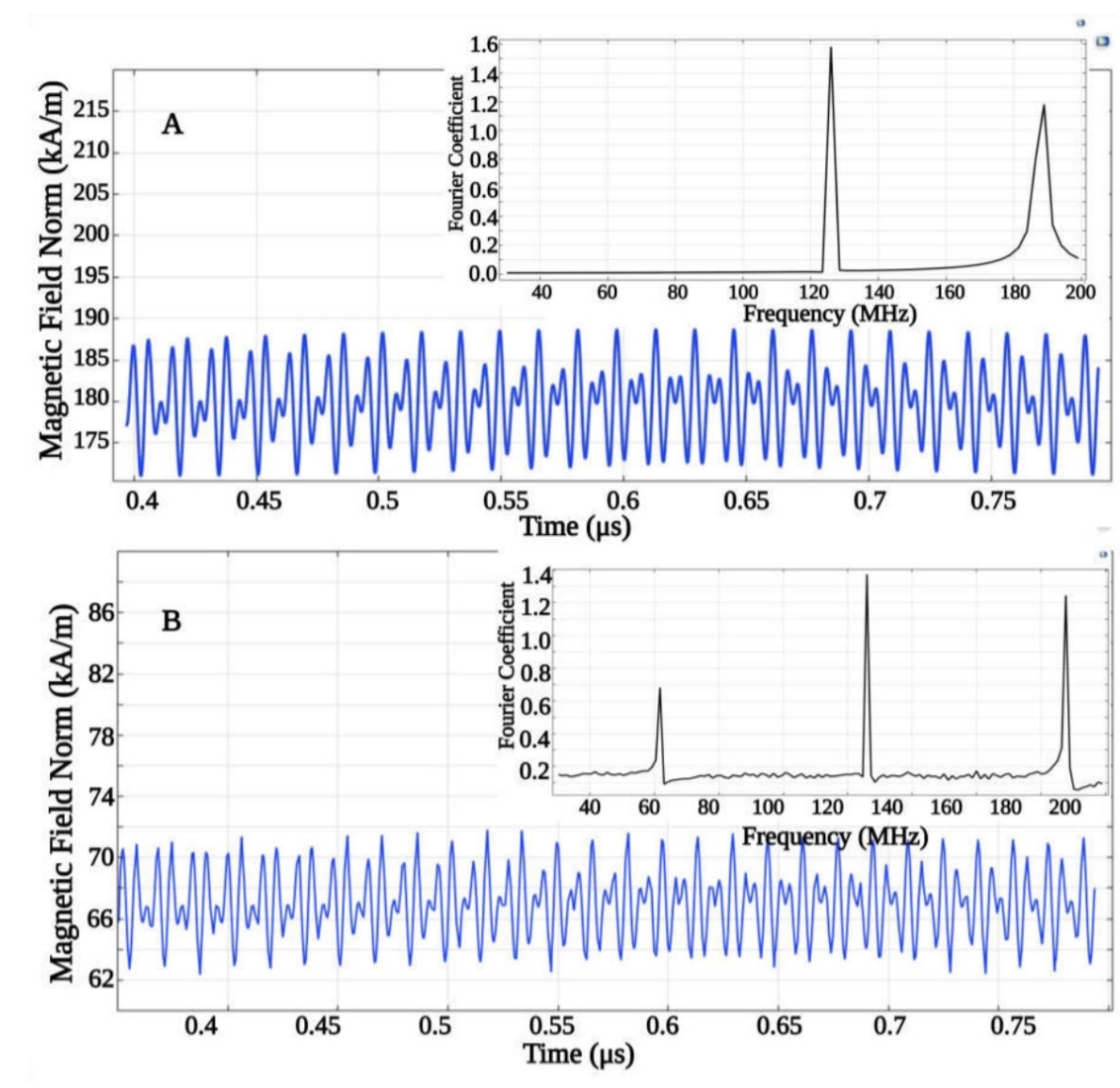

**Figure 4: Normalized magnetic field induced in (A) air medium and, (B) in the magnetostrictive core. Respective insets show the frequency components in the signal. The applied magnetic fields are in the two frequencies $f_1$ and $f_2$, which are 126 MHz and 188 MHz respectively. The ME–MTI demodulated frequency component ($\Delta f = f_2 - f_1$) of 62 MHz is seen in the inset of Figure B.**

## 3 SIMULATION RESULTS

This section presents the validation of the effect of ME–MTI phenomena, along with the estimation of current density distribution in the physiological medium surrounding the ME coreshell device. The simulation process is expedited by selecting two optimal frequencies at 126 and 188 MHz with a considerable frequency difference. The parallel simulations demonstrate that the magnetostrictive device operating in its Eigen mode generates intense electrical hotspots, resulting in increased electrical polarization in the piezo layer. Furthermore, the study observes waveform variation in the ME mediated MTI setup, and the frequency analysis confirms the multiplication of the two waveforms at the difference frequency of 62 MHz. The induced magnetic field on the magnetostrictive core was studied at an applied DC magnetic bias of 293 mT, as shown in Figure 4, and the surface average of the induced magnetic field was calculated to be $\approx 66.7\ kA/m$. The core operation as a frequency demodulator is also depicted in the inset of Figure 4. Additionally, the time domain variation of the surface average magnetic field in both air and the magnetostrictive shell was investigated, and the inlets of Figure 4 show the demodulated frequency components. The ME mediated analysis confirms the demodulation of the interference frequency.

### 3.1 Stimulation current density ($i_{ind}$) analysis in the physiological medium

We anticipate that in practical scenarios, the minimum local distance between the ME microdevice and the intended cell/tissue surface could be in units of $10\mu m$ range. We calculated the surface average of the current density over the physiological spatial frame at intervals of $10\mu m$, until $100\mu m$ from the piezo shell surface. Each of the coloured concentric circles in Figure 5 represents $10\mu m$ thickness of muscle layer. Figure 5 shows the current density dissipation at specific distances from the piezoelectric surface.

Through a parallel analysis, we determine that an optimum current density required for stimulation ranges from 8–18 $\mu A/cm^2$. Below this threshold, we see sporadic generation of membrane potential and above the threshold, hyper–polarization of the cell surface was observed. We estimated the current density ($i_{ind}$) dissipating into the physiological medium at different time instances over the spatial intervals of 20, 30 and 40 $\mu m$, from the piezo shell surface, as shown in Figure 6. We infer that the dissipation current density diminishes exponentially over the dissipation medium and reduces below the needed threshold beyond $25 - 30\ \mu m$.

### 3.2 Bio–stimulation

We used the Hodgkin–Huxley (HH) neuron model [13] to demonstrate the electrical activation of a neuron by evoking an action potential or a spike. The action potential is an elementary, stereotyped impulse generated and exchanged by neurons. In the model, the generation of action potential is controlled mainly through the voltage gated potassium ($K^+$), sodium ($Na^+$), a leaky current channel, and the induced current ($i_{ind}$), which is generated from the non–linear action of the ME coreshell. The governing equation

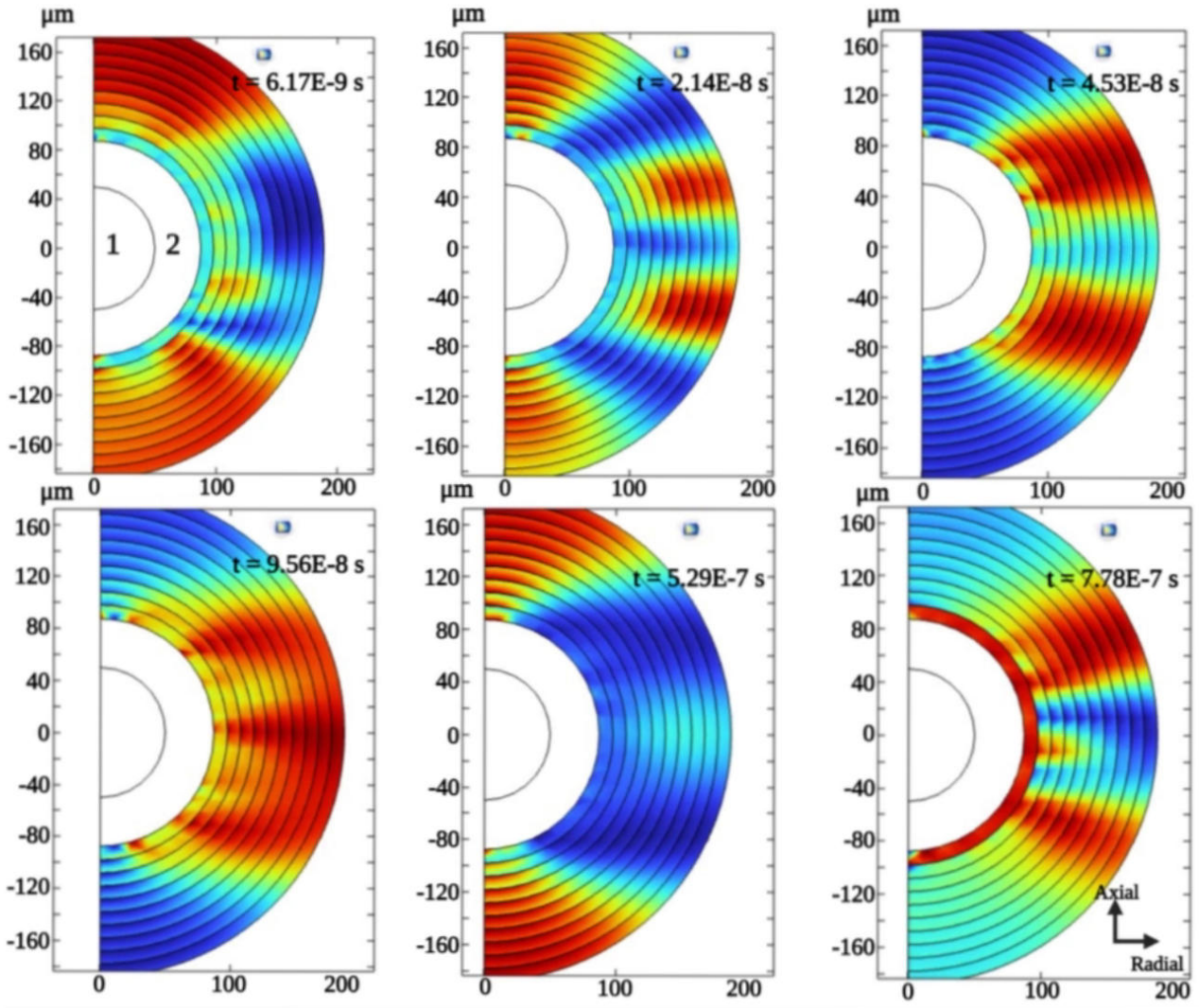

**Figure 5: Dissipation of the current density from the piezo shell surface into the muscle medium. The magnetostrictive core and the piezo shell are numbered as '1' and '2' respectively, and each other concentric circles represent $10\mu m$ muscle layers.**

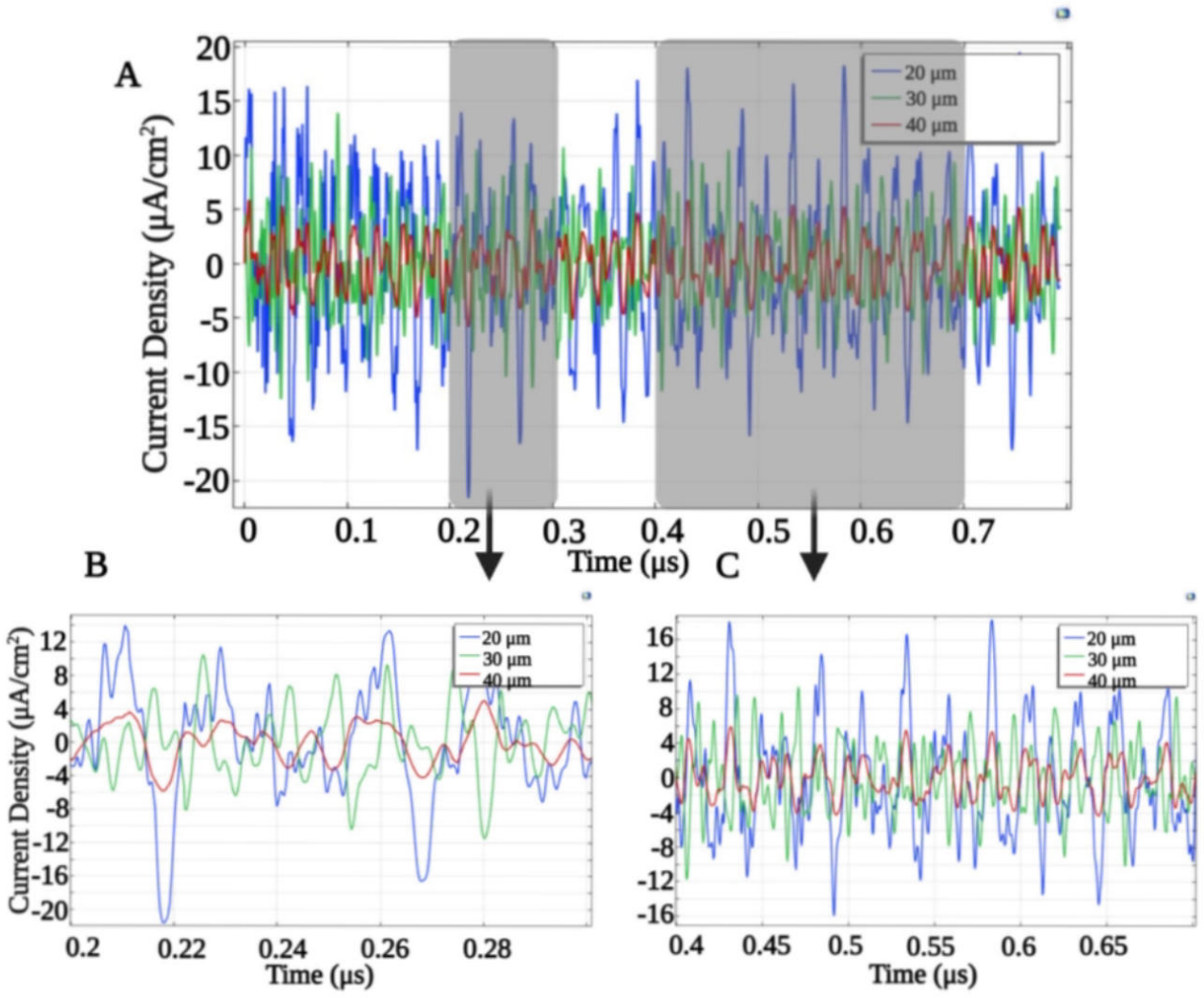

**Figure 6: Current density distribution over the spatial muscle frame for $20\ \mu m$, $30\ \mu m$ and $40\ \mu m$ distance from the piezo shell surface. The distribution density falls below the said threshold of $8\ \mu A/cm^2$ beyond $30\ \mu m$. The shaded area shows specific time periods and the current distribution in those periods (B&C)**

for the generation of the membrane potential is given below:

$$\frac{\mathrm{d}v_m}{\mathrm{d}t} = -\frac{1}{C_m}\Big[g_{\mathrm{K}}\left(v_m - V_{\mathrm{K}}\right) + g_{\mathrm{Na}}\left(v_m - V_{\mathrm{Na}}\right) + + g_L\left(v_m - V_L\right) - \underbrace{i_{ind}}_{\text{stimulation}}\Big], \tag{2}$$

**Table 1: Parameters that quantify the electrical activity of a neuron**

| Parameter | Nominal value | Unit |
|---|---|---|
| Initial membrane conductance, Potassium channel ($\bar{g}_{\mathrm{K}}$) | 36 | $\mathrm{mS/cm^3}$ |
| Nernst potential, Potassium channel ($V_{\mathrm{K}}$) | -70 | mV |
| Initial membrane conductance, Sodium channel ($\bar{g}_{\mathrm{Na}}$) | 120 | $\mathrm{mS/cm^3}$ |
| Nernst potential, Sodium channel ($V_{\mathrm{Na}}$) | 50 | mV |
| Initial membrane conductance, other ion channels ($g_L$) | 0.3 | $\mathrm{mS/cm^3}$ |
| Combined Nernst potential, other ion channels ($V_L$) | -54.4 | mV |
| Specific membrane capacitance ($C_m$) | 1 | $\mu\mathrm{F/cm^2}$ |

where $C_m$ is specific membrane capacitance; $V_{\mathrm{K}}$, $V_{\mathrm{Na}}$, and $V_L$ are Nernst potentials for $K^+$ ions; $Na^+$ and other ions are combined as "leak" channels; and $g_{\mathrm{K}}$, $g_{\mathrm{Na}}$, and $g_L$ are the corresponding membrane conductances. Voltage–gated conductances $g_{\mathrm{K}} = \bar{g}_K m_{\mathrm{K}}{}^4$ and $g_{\mathrm{Na}} = \bar{g}_{\mathrm{Na}} m_{\mathrm{Na}}^3 h_{\mathrm{Na}}$ change with time during an action potential. $m_{\mathrm{K}}{}^4$ and $m_{\mathrm{Na}}^3 h_{\mathrm{Na}}$ represent the opening probabilities for $K^+$ and $Na^+$ channels, respectively. The gating variables $m_{\mathrm{K}}$, $m_{\mathrm{Na}}$, and $h_{\mathrm{Na}}$, steady state equations and the gating functions are fixed values and could be referred here [11] (in Section 1.9 and 1.10 of the cited reference). The general parameters used to compute the membrane potential $v_m$ is given in Table 1.

All the supplementary equations that govern the generation of membrane potential in our model were included from the model here [21]. We used COMSOL Multiphysics, where we included the Global ODEs and DAEs (*ge*) module to input all the parameters and the global equations relating to the HH model. The current density value from our previous ME coreshell simulation was interpolated and included as a piecewise function which was then included in the model equation. Figure 7 shows the generated single spike of the membrane potential. The perturbation frequencies were selected to be in the MHz range to reduce the complexity of the Multiphysics simulations. Due to this selection, the time period of the equivalent current density on the piezo shell is in the $\mu s$ range. Thus, the current density values were time scaled by a factor of $10^4$ to compensate the high frequency selection. The values were then interpolated and the resultant $i_{ind}$ was substituted in Eq. 2 as a piecewise function.

## 4 CONCLUSION

We have demonstrated a proof–of–concept for a novel electronics-free implantable magnetoelectric microdevice, capable of stimulating cells and tissues in the body through a multifunctional hybrid composite nature. Unlike traditional methods, the device is self contained, battery- free, and controllable, which eliminates the need for electronic circuits and wires. The use case scenario presented in this work showcases the device's ability to function as a frequency demodulator and create a localized, tunable, and stirrable electric potential on the piezoelectric shell, triggering electrical spikes in neural cells. Our results indicate that the ME coreshell can be a potential candidate for wireless and non–invasive cellular stimulation therapy with high resolution and precision. These findings pave the way for developing injectable material structures for effective cellular stimulation, enabling a range of biomedical applications in the future.

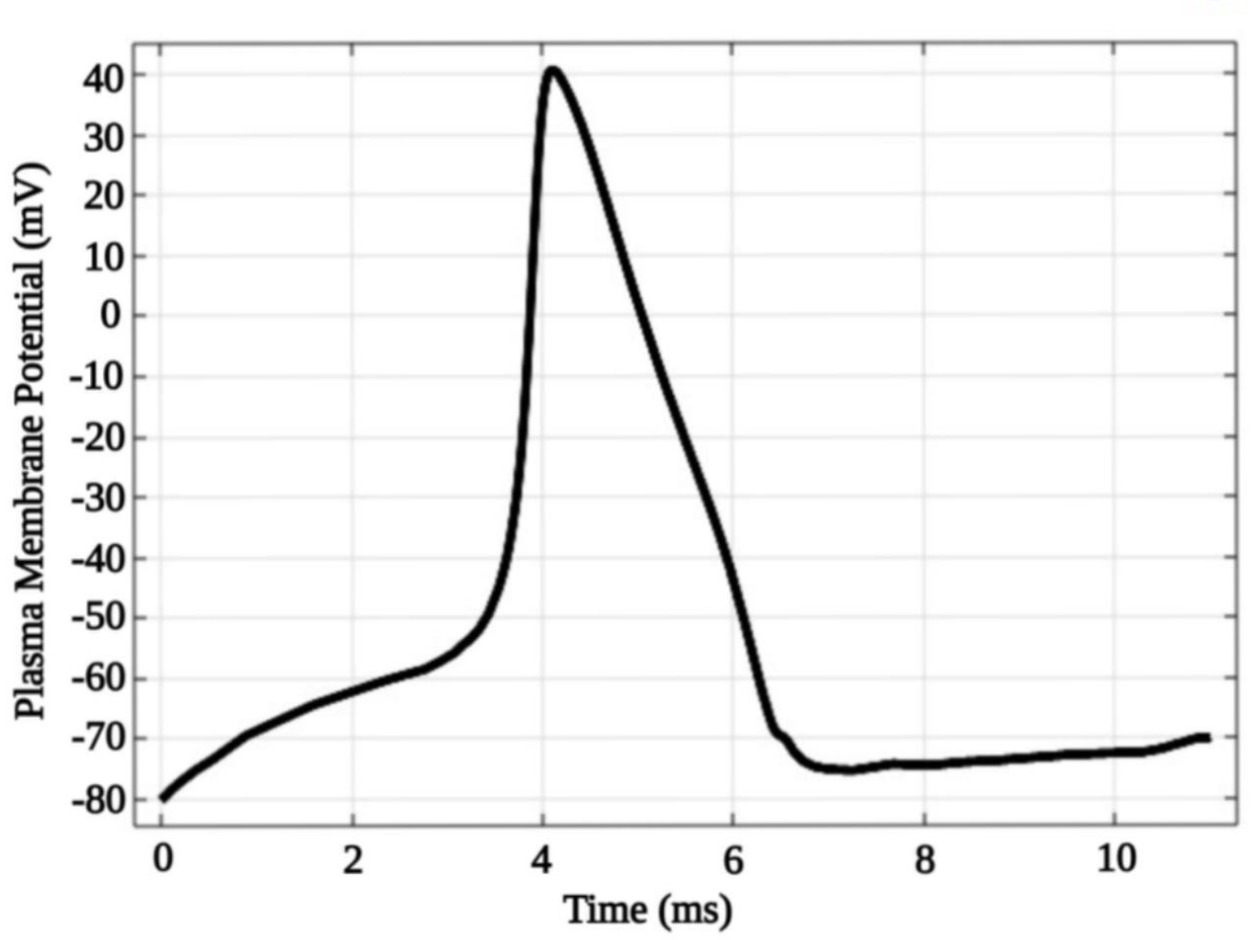

**Figure 7: Generated membrane potential through the induced stimulation current $i_{ind}$. The repolarization and depolarization phases of the signal are in accordance with the membrane potential shown in the reference here [21] (Page 32, Figure 4.1).**

## ACKNOWLEDGMENTS

The work has been supported by the project Brain–Connected inteRfAce TO machineS (B–CRATOS) under grant #965044, Horizon 2020 FET–OPEN. This work was carried out during the tenure of the ERCIM 'Alain Bensoussan' Fellowship Programme awarded to the first author. All figures were either created or edited with BioRender.com.